\documentclass[osa,twocolumn,showpacs,superscriptaddress]{revtex4-1}
\usepackage{xcolor}
\usepackage{graphicx}
\usepackage{subfigure}
\usepackage{textcomp}       
\usepackage{amsmath,amssymb}
\usepackage[colorlinks=true, citecolor=blue, linkcolor=red]{hyperref}
\usepackage{url}

\newcommand{\microrad}{\textmu{}rad }

\begin{document}

\newcommand{\R}[1]{\textcolor{red}{#1}}

\title{Angular control of
  optical cavities in a radiation pressure dominated regime: the
  Enhanced LIGO case}

\author{Katherine L Dooley}
\email[Corresponding author: ]{katherine.dooley@ligo.org}
\altaffiliation[Current address: ]{Albert-Einstein-Institut, Max-Planck-Institut f\"{u}r Gravitationsphysik, D-30167 Hannover, Germany}
\affiliation{University of Florida, PO Box 118440, Gainesville, FL 32611, USA}

\author{Lisa Barsotti}
\affiliation{LIGO Laboratory, Massachusetts Institute of Technology, Cambridge, MA 02139, USA}

\author{Rana~X~Adhikari}
\affiliation{LIGO Laboratory, Division of Physics, Math, and Astronomy, California Institute of Technology, Pasadena, CA 91125, USA}

\author{Matthew~Evans}
\affiliation{LIGO Laboratory, Massachusetts Institute of Technology, Cambridge, MA 02139, USA}

\author{Tobin~T~Fricke}
\altaffiliation[Current address: ]{Albert-Einstein-Institut, Max-Planck-Institut f\"{u}r Gravitationsphysik, D-30167 Hannover, Germany}
\affiliation{Louisiana State University, Baton Rouge, LA 70803, USA}

\author{Peter~Fritschel}
\affiliation{LIGO Laboratory, Massachusetts Institute of Technology, Cambridge, MA 02139, USA}

\author{Valera~Frolov}
\affiliation{LIGO - Livingston Observatory, Livingston, LA 70754, USA}

\author{Keita Kawabe}
\affiliation{LIGO - Hanford Observatory, Richland, WA 99354, USA}

\author{Nicol\'as Smith-Lefebvre}
\altaffiliation[Current address: ]{LIGO Laboratory, Division of Physics, Math, and Astronomy, California Institute of Technology, Pasadena, CA 91125, USA}
\affiliation{LIGO Laboratory, Massachusetts Institute of Technology, Cambridge, MA 02139, USA}

\begin{abstract}
We describe the angular sensing and control of the 4\,km detectors of
the Laser Interferometer Gravitational-wave Observatory (LIGO). The
culmination of first generation LIGO detectors, Enhanced LIGO operated
between 2009 and 2010 with about 40 kW of laser power in the arm
cavities. In this regime, radiation pressure effects are significant
and induce instabilities in the angular opto-mechanical transfer
functions. Here we present and motivate the angular sensing and
control (ASC) design in this extreme case and present the results of
its implementation in Enhanced LIGO. Highlights of the ASC performance
are: successful control of opto-mechanical torsional modes, relative
mirror motions of $\le 1\times 10^{-7}$ rad rms, and limited impact on
in-band strain sensitivity.
\end{abstract}

\maketitle

\newpage


\section{Introduction}
Over the last decade, a world-wide network of ground based laser
interferometers~\cite{Adhikari2013Gravitational} has been constructed
and operated in pursuit of the first direct detection of gravitational
waves (GWs). The U.S. Laser Interferometer Gravitational-wave Observatory
(LIGO)~\cite{Abbott2009LIGO} operates detectors in Livingston, LA and
Hanford, WA, each consisting of a suspended Michelson interferometer
with 4\,km Fabry-Perot arm cavities. These detectors attained their
best sensitivity yet during the most recent scientific data taking
run, known as ``S6'', which took place between July 2009 and October
2010, in a configuration called ``Enhanced
LIGO''~\cite{Adhikari2006Enhanced, Frede2007Fundamental,
  Dooley2012Thermal, Fricke2012DC}.  Enhanced LIGO featured several
improvements with respect to the earlier Initial LIGO configuration
(2001-2007). One of the critical upgrades was the increase in the
laser power circulating inside the arm cavities by about a factor
four. The 40\,kW of laser power stored in the Enhanced LIGO cavities
greatly complicated the relative alignment of the interferometer
mirrors.  For the laser interferometer to operate properly, its
mirrors must be aligned to each other with a relative rms misalignment
not larger than about a tenth of a microradian. Meeting this stringent
requirement is particularly challenging in the presence of radiation
pressure effects.

Radiation pressure exerts torque on the suspended mirrors, adding to
the fixed restoring torque of the suspension. The possibility of this
torque to de-stabilize optical cavities was first recognized in 1991
by Solimeno et al.~\cite{Solimeno1991FabryPerot}. By 2003, it was
clear in the LIGO community that the effect of radiation pressure on
angular dynamics was relevant for LIGO~\cite{Sidles2003Optical} and
the full details of the effects were described by Sidles and Sigg in
2006~\cite{Sidles2006Optical}. Fan et al. measured the predicted
optical-mechanical torsional stiffness at the Gingin Facility in
Australia \cite{Fan2009Observation}, Driggers et. al. demonstrated its
effect at the Caltech 40m prototype~\cite{Driggers2006Optomechanical}
and Hirose et al. showed that although the optical torque in Initial
LIGO (about 10\,kW of laser power circulating in the Initial LIGO arm
cavities) was measurable and similar in magnitude to the suspension
restoring torque, it was not yet significant enough to require a
change to the angular controls~\cite{Hirose2010Angular}. In this
paper we show the effect of optical torque in the Enhanced LIGO
interferometers and also present the design concept and implementation
of an alignment sensing and control scheme (ASC) which allowed us to
operate an interferometer with angular mechanics dominated by
radiation pressure.

Two of the authors (Barsotti and Evans) created a numerical model of
the ASC for Enhanced LIGO that specifically included radiation
pressure torque~\cite{Barsotti2009Modeling}. They showed that, in
principle, the radiation pressure torque can be controlled without
detrimental consequences to the sensitivity of the detector. The
proposed solution rotates the control basis to one that naturally
represents the eigenmodes of mirror motions coupled by radiation
pressure. We implemented this control scheme on the Enhanced LIGO
interferometers with up to 40\,kW of circulating power, successfully
controlling the angular degrees of freedom in the presence of the
radiation pressure instability. The demonstrated solution meets the
LIGO requirements and is extensible to the next generation of LIGO
detectors currently under construction, Advanced LIGO.

\begin{figure*} 
\centering
  \includegraphics[width=16cm]{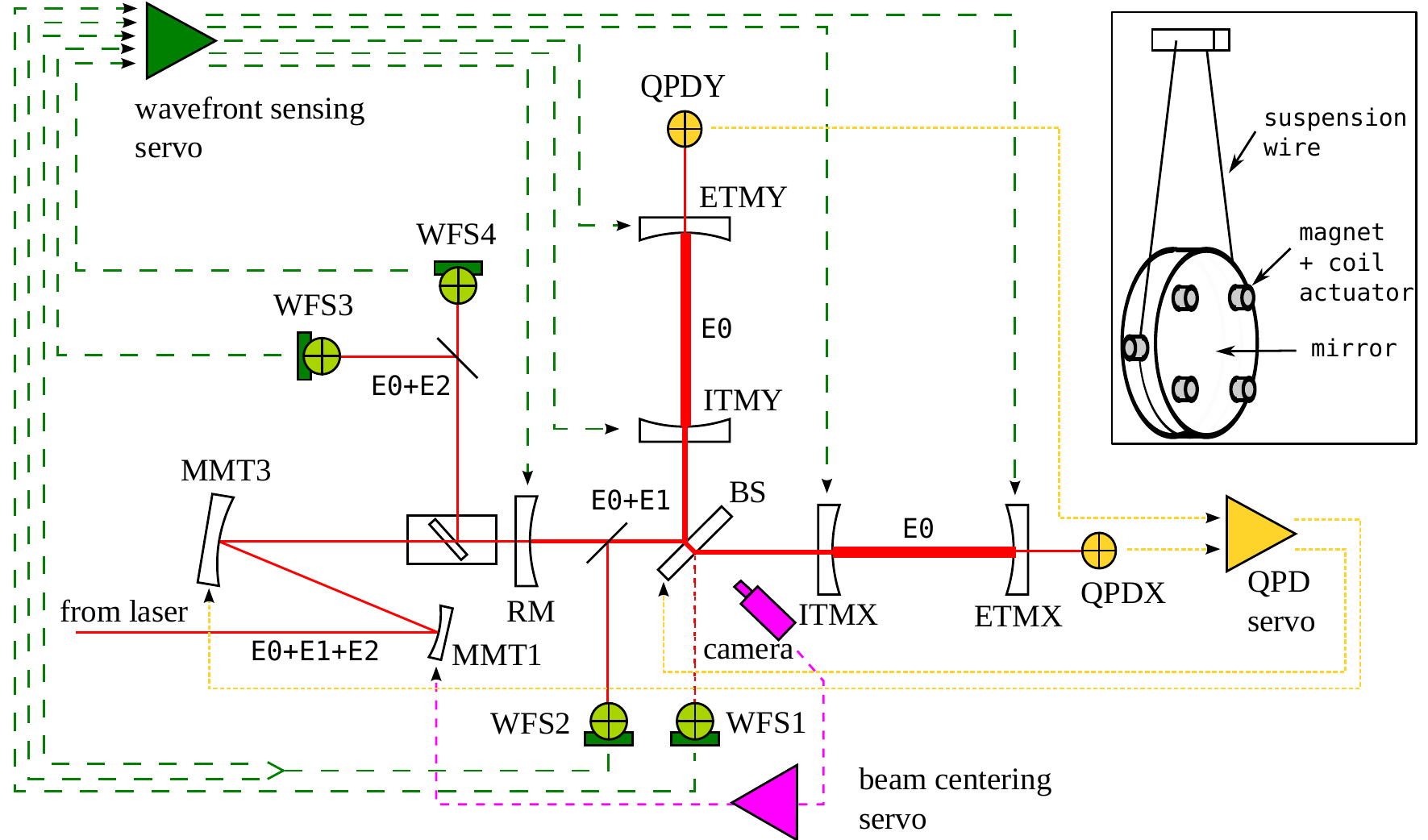}
  \caption{Power-recycled Fabry-Perot Michelson interferometer layout,
    with ASC system superimposed. The 8 actively aligned mirrors
    (ITMs, ETMs, MMTs, BS, and RM) and the ASC sensors (WFS, QPDs, and
    camera) are shown. All additional optics are omitted for
    simplicity. The QPD and beam centering servos provide drift
    control on minute time scales. The wavefront sensing (WFS) servo
    maintains the alignment of the interferometer mirrors with respect
    to each other up to several Hz. Both I and Q phases are used from
    WFS2, whereas only one quadrature is read out from each of the
    other WFS. The carrier field is $E0$ and sideband fields $E1$ and
    $E2$ are respectively resonant and non-resonant in the power
    recycling cavity. All test masses are suspended as single stage
    pendula as depicted in the upper right corner, and are outfitted
    with magnet-coil actuators to control angular and longitudinal
    degrees of freedom.}
\label{fig:ASC}
\end{figure*}

The interferometer layout and the control scheme are introduced in
Section~\ref{sec:alignment}. Section~\ref{sec:design} presents the
modified design after a review of the physics of radiation pressure
induced torque on the mirrors. This section also highlights a direct
measurement of the opto-mechanical modes that are
controlled. Section~\ref{sec:results} presents the results of using
the new alignment control scheme at high laser powers, including the
residual mirror motion and the noise performance. Key differences and
implications for Advanced LIGO are outlined in
Section~\ref{sec:aLIGOASC}, and Section~\ref{sec:summary} provides a
summary. All data presented are from the Livingston Observatory;
results from the Hanford Observatory are similar.

\section{Experimental setup} 
\label{sec:alignment}

Each LIGO detector is a power-recycled Fabry-Perot
Michelson laser interferometer featuring suspended test masses
(mirrors) in vacuum. A stabilized laser beam (with a wavelength of
1064\,nm) is directed to the interferometer, whose two arm lengths are
set to maintain nearly destructive interference of the recombined
light at the Michelson (dark) anti-symmetric port. An appropriately
polarized GW differentially changes the arm lengths,
producing a signal at the anti-symmetric port proportional to the
GW strain.  The test masses are suspended by a single
loop of steel wire to provide isolation from ground motion, as
depicted in Fig.~\ref{fig:ASC}. Each mirror is equipped with five
magnet-coil actuators to control the mirror's longitudinal and angular
position. Furthermore, the carrier laser field is phase modulated by
an electro-optic modulator at 24.4\,MHz and 61.1\,MHz to generate
sidebands for use in a modulation-demodulation technique of sensing
the interferometer's longitudinal and angular degrees of freedom.

There are several reasons why the interferometer's mirrors must be
actively aligned:
\begin{itemize}
\item to maximize optical power coupling \vspace{-5pt}
\item to suppress motion from external disturbances \vspace{-5pt}
\item to counteract a static instability at high laser power
\end{itemize}
The requirements for how much residual motion is
tolerable~\cite{Fritschel1997Alignment, Fritschel1998Alignment} stem
from the mechanisms by which misalignment couples to strain
sensitivity. The most significant coupling of angular motion to cavity
length occurs when the beam spot is off-center from the mirror's axis
of rotation. The combination of mirror angular motion
$\theta_{\mathrm{mirror}}(f)$ and beam spot motion on the test masses
$d_{\mathrm{spot}}(f)$ changes the length of the arms by:
\begin{equation}
\Delta L(f) = d_{\mathrm{spot}}^{\mathrm{RMS}} \times \theta_{\mathrm{mirror}}(f) + \theta_{\mathrm{mirror}}^{\mathrm{RMS}} \times d_{\mathrm{spot}}(f)
\label{eq:A2L}
\end{equation}
and results in an increase in the sensed longitudinal motion. The
relevant quantities for describing the mirror's motion are its
root-mean-square (rms) and in-band (audio frequency) noise. It is
worth noting that once all of the interferometer cavities are brought
to resonance and the DC pointing no longer contributes to the rms, the
rms is dominated by the pendular motion.

There are additional mechanisms by which misalignment affects
displacement sensitivity. First, a high order effect arises because
misalignments affect power build-up quadratically which in turn
modulates the noise floor in the shot-noise-limited regime. A second
mechanism results as a side effect of having active angular
alignment. Due to imperfections in the actuators, there will always be
a small amount of longitudinal acutation along with the desired
angular actuation.

\begin{figure} 
\centering 
\includegraphics[width=\columnwidth]{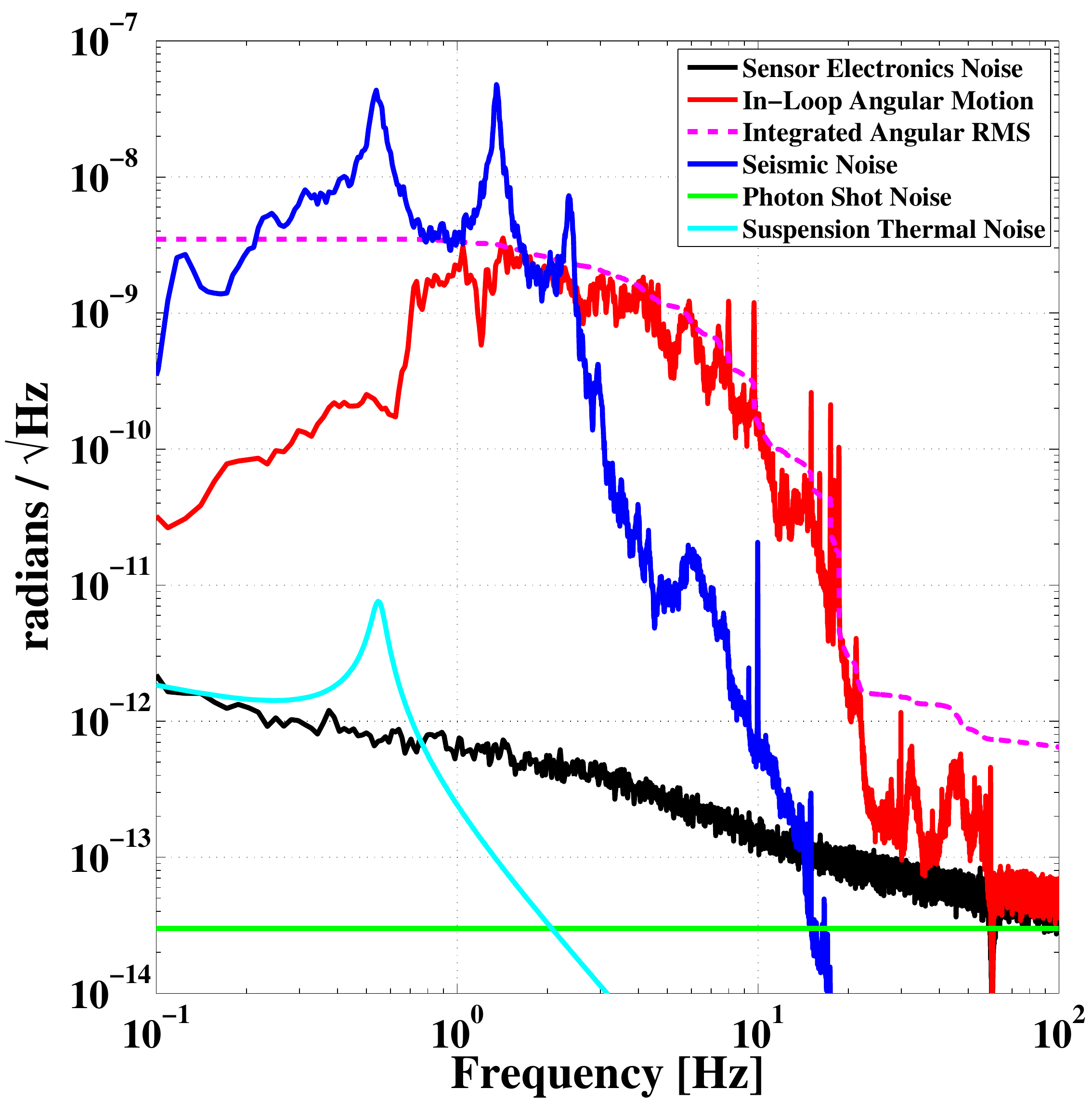}
\caption{Noise budget of an alignment sensor (WFS1) for the pitch
  degree of freedom. Curves for seismic noise, suspension thermal
  noise, and sensor noise are shown. Note that direct seismic and
  suspension thermal noises are small in the GW band, and the largest
  motions are impressed by our control system. Sensing noise dominates
  above approximately 20\,Hz where the seismic isolation stacks
  strongly isolate.}
\label{fig:WFS_NB}
\end{figure}

External disturbances that cause misalignment include: seismic noise,
pitch/yaw mode thermal noise, length-to-angle coupling, acoustic
noise, and radiation pressure torque. Mechanical and electrical design
of suspensions and sensors, isolation in vacuum, and periodic
balancing of mirror actuators are measures taken to reduce the level
of angular motion in the first place. An active control system is used
to mediate the motion that remains, which in turn is itself a source
of misalignments due to sensing noise. As reflected in the noise
budget of one of the alignment sensors in Figure~\ref{fig:WFS_NB},
direct seismic and suspension thermal noises are in fact quite
small. Above 20\,Hz where the seismic isolation platforms strongly
isolate, sensor noise dominates. As a result, the angular motions of
the cavities above these frequencies are dominated by the control
system itself. Sensor noise is thus a primary consideration in servo
design.

The alignment of the interferometer is accomplished via feedback and
there are several frames of reference to which the mirrors are
aligned. Ultimately, the mirrors must be aligned to one another, and
this will be presented in detail shortly. Each individual
optic also has two servos of its own to provide velocity damping.  First,
local shadow sensors provide damping around the pitch and yaw
eigenfrequencies of the mirrors (0.6\,Hz and 0.5\,Hz,
respectively). This damping is relative to the suspension cage which
is already isolated at high frequencies. Second, optical levers
mounted to heavy piers on the ground provide a reference to the local
ground motion. They are more sensitive than the shadow sensors and
serve to suppress the motion which arises from the isolation table
stack resonances from 0.2\,Hz to 2\,Hz. The interaction of these two
velocity damping servos with the main alignment servo results in some
increased complexity of the main servo design.

The fundamental physical principle behind sensing relative mirror
misalignment is the fact that when an optical cavity is misaligned
relative to an incident field, a $\mathrm{TEM}_{01}$ Hermite-Gaussian
mode is generated with an amplitude proportional to the misalignment
\cite{Anderson1984Alignment}. Alignment signals are produced by
directing some of this light onto a quadrant photodiode (QPD), where
the interference of the $\mathrm{TEM}_{00}$ fundamental mode and
$\mathrm{TEM}_{01}$ misalignment mode at the sideband frequency can be
compared on each half of the split diode. The QPD together with the
resonant RF circuit and demodulation system is called a wavefront
sensor (WFS). The amplitude of the alignment signal is a function of
the relative Gouy phase \cite{Siegman1986Lasers} between the
$\mathrm{TEM}_{00}$ and $\mathrm{TEM}_{01}$ modes, which is a function
of the longitudinal position of the detector along the optical
axis. Angular misalignments of different combinations of mirrors can
therefore be distinguished by placing detectors at different locations
along the optical path. The basic formalism of how alignment signals
are generated is presented in Ref.~\cite{Hefetz1997Principles,
  Sampas1990Stabilization, Morrison1994Automatic}.

A detailed description of the control scheme design for the Initial
LIGO configuration is found in Ref.~\cite{Fritschel1998Alignment} and
key aspects relevant for the description of the Enhanced LIGO ASC are
provided here. There are 8 mirrors whose pitch (rotation about the
mirror's horizontal axis) and yaw (rotation about the vertical axis)
angles must be sensed and controlled. The sensing is accomplished
through the use of 8 sensors, which can be organized into three types:
\begin{itemize}
\item wavefront sensors (WFS1, WFS2\footnote{Two signals are derived
  from WFS2.}, WFS3, WFS4) which sense the angular misalignment of the
  cavities with respect to their input beams \vspace{-7pt}
\item CCD image of the beam spot on the beam splitter (BS) \vspace{-7pt}
\item quadrant photodiodes (QPDX, QPDY) which see the beam transmitted
  through the arm cavities
\end{itemize}

Figure~\ref{fig:ASC} shows the basic power-recycled Michelson
interferometer layout, highlighting the locations of these angular
sensors and the eight mirrors they must control. Two WFS, separated in
Gouy phase, are located at the reflected port of the interferometer
where common mode signals appear. The third sees a pick-off of light
from the recycling cavity which contains common and differential
signals, and the fourth gets a pick-off of the light at the
anti-symmetric port where differential mode signals are
transmitted. The common mode represents motion where the optics of one
arm rotate in the same direction as those in the other arm and
differential represents rotations in opposite directions. The eight
mirrors include the four test masses that make up the Fabry-Perot arm
cavities (ITMX, ITMY, ETMX, ETMY), the beam splitter (BS), the
recycling mirror (RM), and two input beam directing mirrors that also
serve as a mode matching telescope (MMT1 and MMT3).

The CCD image and the QPDs are used in slow feedback loops as part of
drift control servos to maintain the beam spot positions at the three
corners of the interferometer. Their bandwidths are below a few~mHz
and below 0.1\,Hz, respectively, and are significantly lower than the
bandwidths of the WFS loops, which keep the mirrors aligned to one
another from DC up to several Hertz.

\begin{figure}
\centering
\includegraphics[width=\columnwidth]{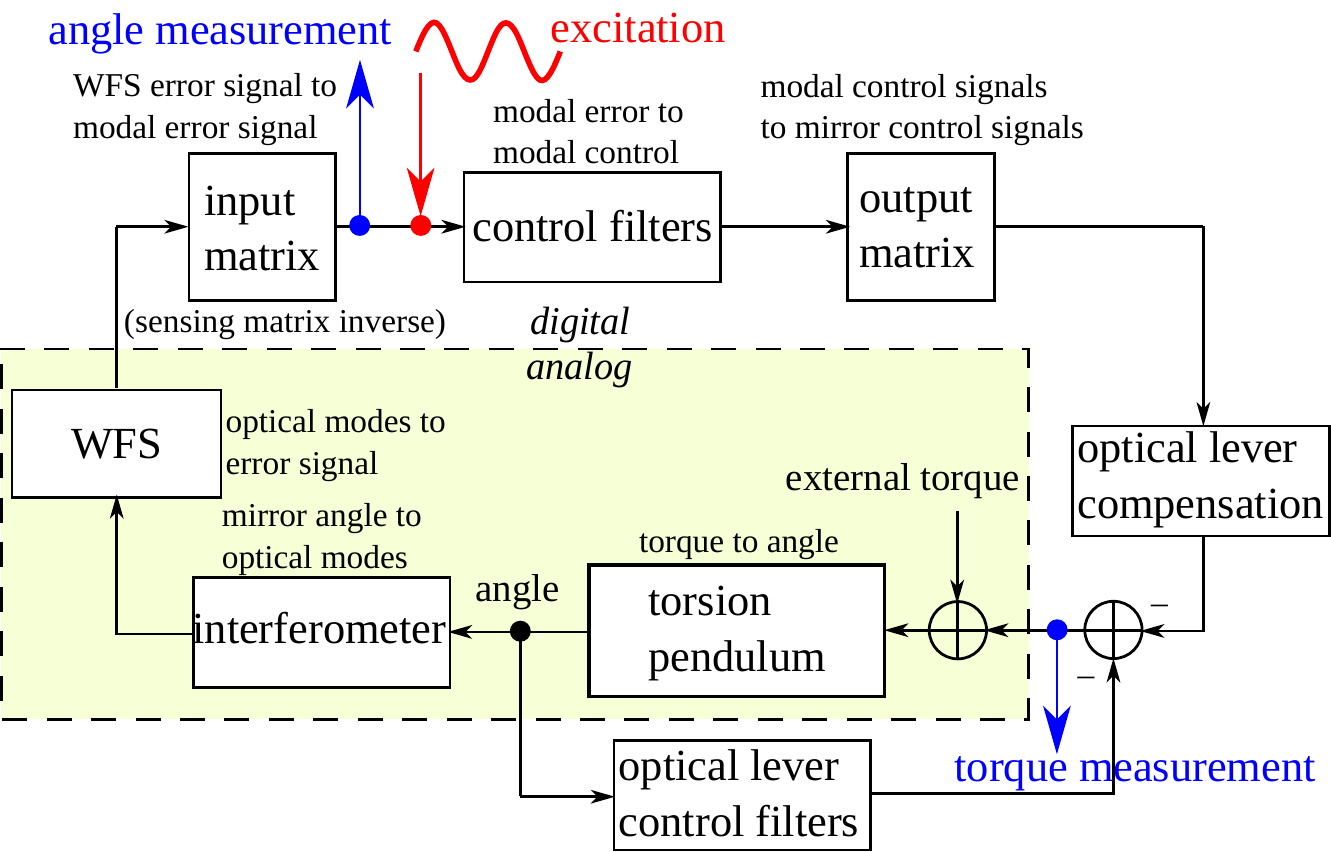}
\caption{Block diagram of major components of the angular control
  servo. The input matrix is the inverse of the sensing matrix
  presented in Table~\ref{table:sensing}. Components within the dashed
  box are analog.}
\label{fig:block}
\end{figure}

Figure~\ref{fig:block} shows a simplified block diagram of the WFS
servo. The interferometer converts individual mirror motions into
optical modes which in turn are converted into error signals by the
WFS. The angular error signals are digitized, filtered, and converted
into analog control signals for individual mirrors. Two matrices in
series rotate the alignment signals from the WFS basis to the optic
basis. Control filters are implemented in the intermediate basis.

In Initial LIGO, the sensing basis was that of common and differential
ETM/ITM motion and the RM, and servos were designed in this basis. The
input matrix was diagonal and the output matrix was created to send
equal or equal and opposite signals to the ETMs and ITMs,
respectively. In this work, we describe a change of basis to improve
the stability of the interferometer in the presence of radiation
pressure torque.

\section{ASC design in the presence of a radiation pressure instability}
\label{sec:design}

The effectiveness of the Initial LIGO ASC design is limited in the
regime of high circulating power where radiation pressure modifies the
simple pendulum plant in a way which is power-dependent. As is
detailed in this section, torque due to radiation pressure couples the
angular motions of the arm cavity mirrors such that the simple single
resonance of a given mirror's torque-to-angle transfer function splits
into two, with frequency shifts dependent on power. Controlling this
new plant could be accomplished with the Initial LIGO system by
increasing the gains of the WFS loops, but it would be at the expense
of introducing too much control noise in the GW measurement band. An
alternative solution is thus required to achieve both adequate angular
control and minimal noise impression. In this section, we first review
the formalism of radiation pressure torque in cavities. Then, we
present a direct measurement of the opto-mechanical modes of the
Enhanced LIGO arm cavities for several powers. Finally, we describe
the modified control scheme and present its implementation.

\subsection{Torque induced by radiation pressure}
\label{sec:rp}
In the limit of no circulating power in a suspended Fabry-Perot
cavity, each of the individual mirrors has independent equations of
motion.  With power circulating in the cavity, however, radiation
pressure effects couple the equations of motion of the two mirrors.
As a beam impinging a mirror off-center creates a torque, an
opto-mechanical angular spring is created due to the geometric
relationship of beam displacements and mirror angles
\cite{Siegman1986Lasers}.  This fact has two important consequences:
on one hand, as the torque induced by radiation pressure is
proportional to the power stored inside the cavity, the
opto-mechanical angular transfer functions of the cavity mirrors
change as a function of the stored power. On the other hand, for large
powers, radiation pressure can even overcome the restoring torque of
the mirror suspension, creating an unstable system.

To understand how the cavity dynamics are affected by radiation
pressure, it is useful to diagonalize the coupled equations of the
mirror motion into two normal cavity modes.  We refer to
Ref.~\cite{Sidles2006Optical} for a complete derivation of the
torsional stiffness matrix which couples the static misalignment of
the two cavity mirrors, and here we use only the final expressions for
the two eigenvalues $k_{S,H}$ and eigenvectors $v_{S,H}$ of that
matrix: 
\begin{eqnarray}
k_{S,H} & = k_0 \frac{(g_1 + g_2)\pm\sqrt{ (g_1-g_2)^2+4}}{2} \\
v_S & = [1, \frac{k_0}{k_S-k_0 g_1}] \\
v_H & = [\frac{k_0}{k_0 g_2-k_H}, 1]
\label{eq:kSH}
\end{eqnarray}
where $k_0 = \frac{2PL}{c (g_1 g_2 - 1)}$ ($L=3995$ m is the cavity
length, $c$ the speed of light, and $g_1$ and $g_2$ the geometric
$g$-factors of the cavity).

The resonant frequency of each of the opto-mechanical modes can then
be written as:
\begin{eqnarray}
f_{S,H} = \frac{1}{2\pi}\sqrt{\frac{k_{p} + k_{S,H}}{I}}
\end{eqnarray}
where $I$ is the mirror moment of inertia ($I$=0.0507 kg m$^2$), and
$k_{p}$ is the restoring torque of the mirror suspension ($k_{p}=$.72
Nm/rad pitch and 0.5 Nm/rad yaw). For the Initial and Enhanced
LIGO interferometers, the $g$-factors of the cavities are:
\begin{eqnarray}
g_1 &= g_{\mathrm{ITM}} = 1 - L/R_{\mathrm{ITM}} &= 0.726\\
g_2 &= g_{\mathrm{ETM}} = 1 - L/R_{\mathrm{ETM}} &= 0.460    
\end{eqnarray}
so $k_S$ is negative and $k_H$ is positive. 

Known as the Sidles-Sigg effect, the radiation pressure torque either
softens or stiffens the mechanical springs. We therefore refer to the
two modes as ``soft (S)'' or ``hard (H)''.  As power increases, the
frequency of the hard mode increases, but the frequency of the soft
mode decreases until $k_S + k_p < 0$ when there is no longer a real
resonant frequency, corresponding to an unstable system.

\begin{table}
  \centering
  \caption{Resonant frequencies (pitch) in Hz for the soft and hard
    opto-mechanical modes of a typical Initial LIGO circulating power
    (9 kW) and the highest of Enhanced LIGO powers (40 kW). The soft
    mode in Enhanced LIGO is unstable.}
\begin{tabular}{l l l l l}
\hline
& $P_{\mathrm{circ}}$ [kW] & $f_{\mathrm{p}}$ [Hz] & $f_{\mathrm{S}}$ [Hz] & $f_{\mathrm{H}}$ [Hz]\\ 
\hline
Initial LIGO  &   9 & 0.60 &  0.19    & 0.66 \\ 
Enhanced LIGO &  40 & 0.60 & -1.04 & 0.83 \\ 
\hline
\end{tabular}
   \label{table:k_tot}
\end{table}

The values of the opto-mechanical frequencies of the soft and hard
modes for Initial and Enhanced LIGO powers are outlined in
Table~\ref{table:k_tot}. The increase of stored power from 9\,kW in
Initial LIGO to 40\,kW in Enhanced LIGO makes radiation pressure
torques cross into the realm of significance. The torque due to
radiation pressure surpasses the suspension restoring torque such that
the soft opto-mechanical mode, which had just approached instability
in Initial LIGO, actually becomes unstable in Enhanced LIGO.

\subsection{Measurement of opto-mechanical modes}
\begin{figure*}
\centering
\includegraphics[width=1.0\columnwidth]{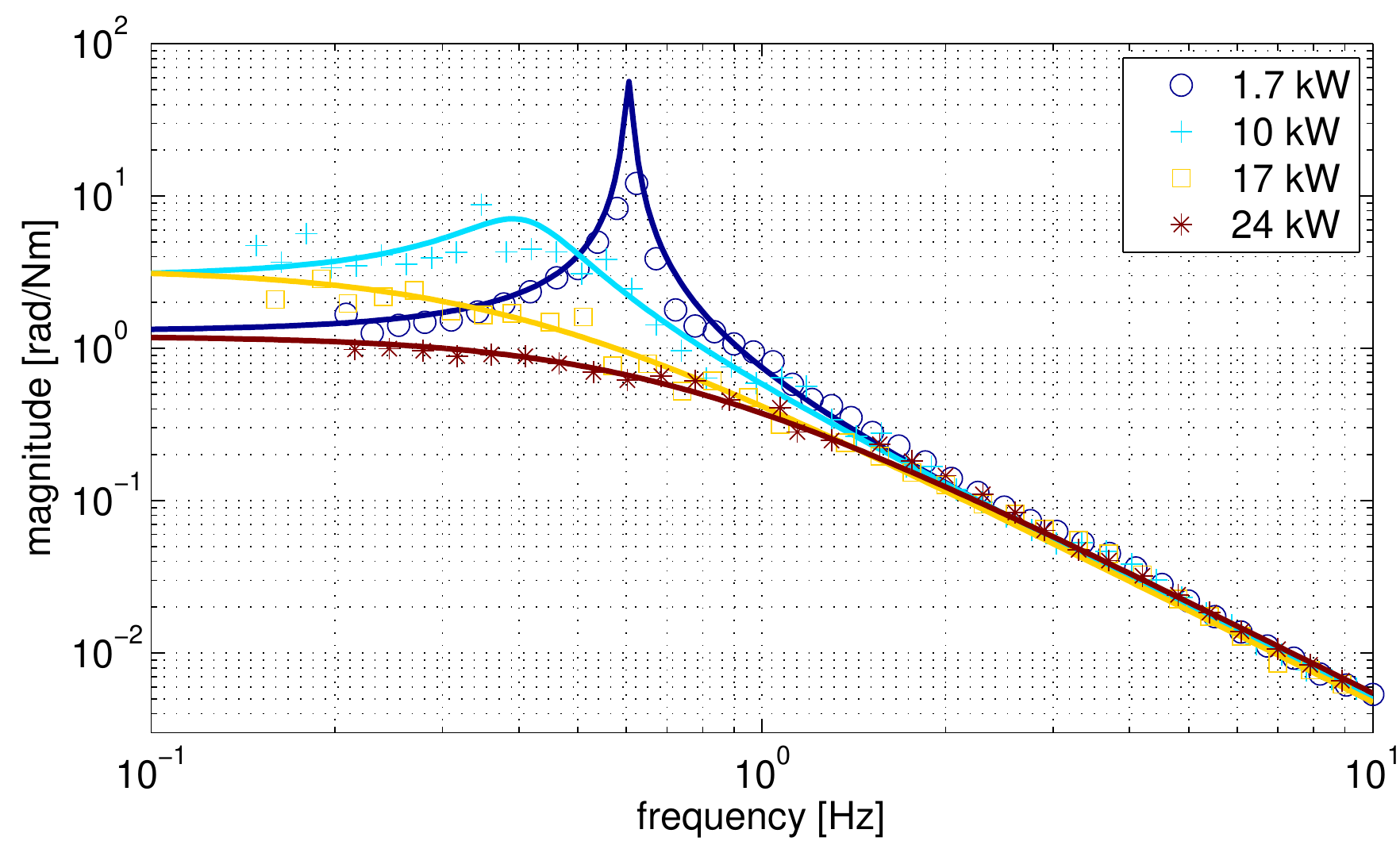}\includegraphics[width=1.0\columnwidth]{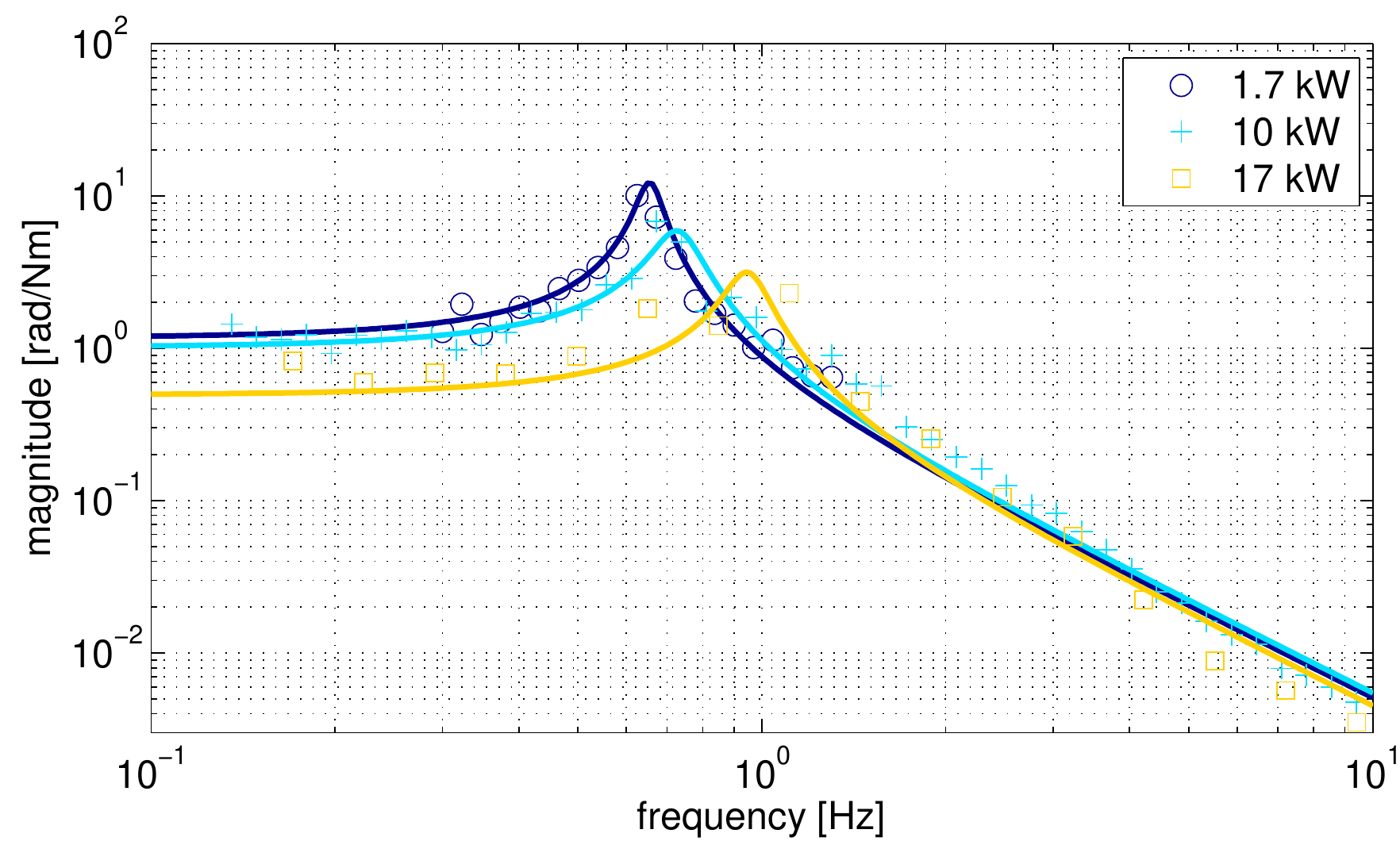}
\subfigure[]{\includegraphics[width=1.0\columnwidth]{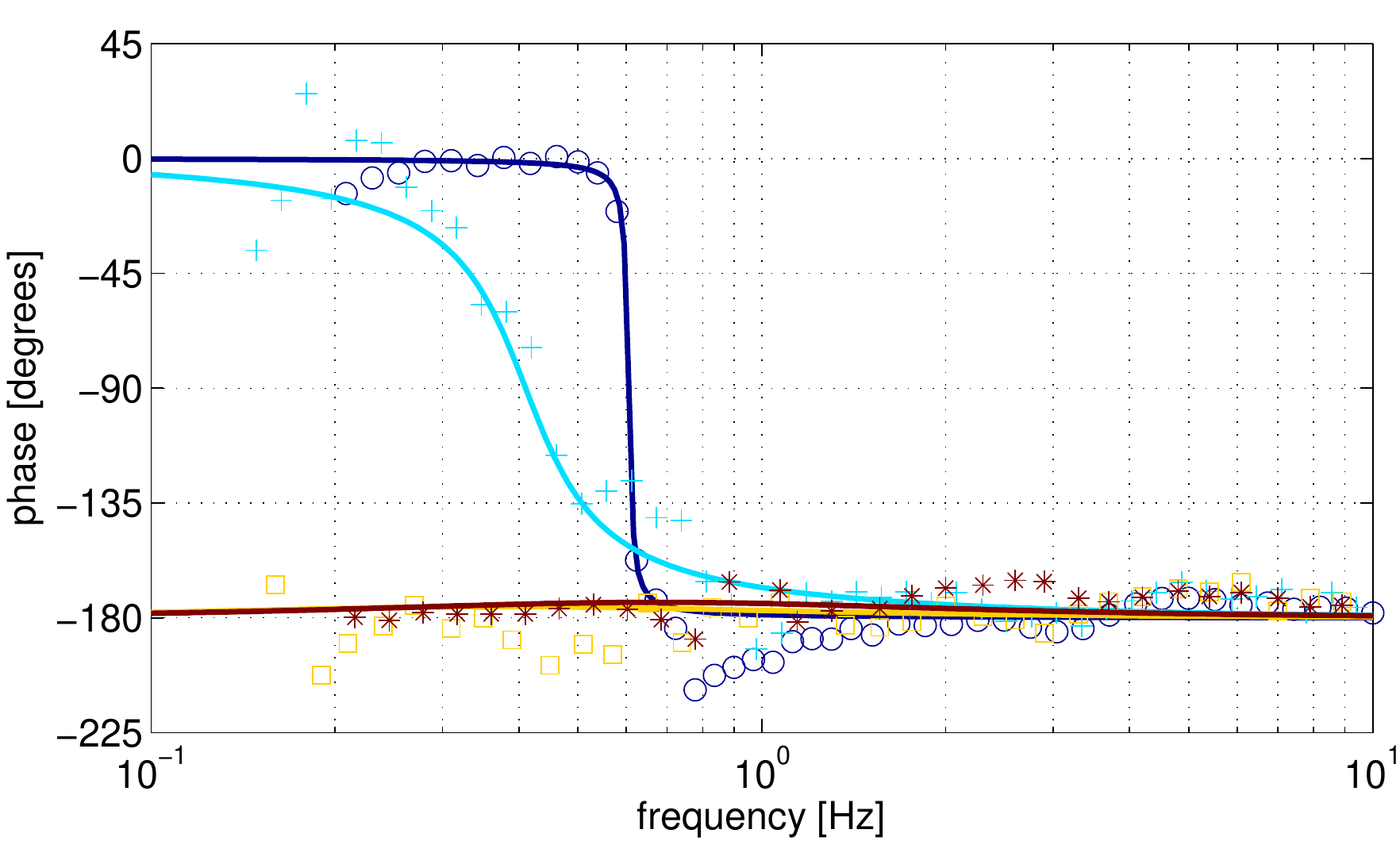}}\subfigure[]{\includegraphics[width=1.0\columnwidth]{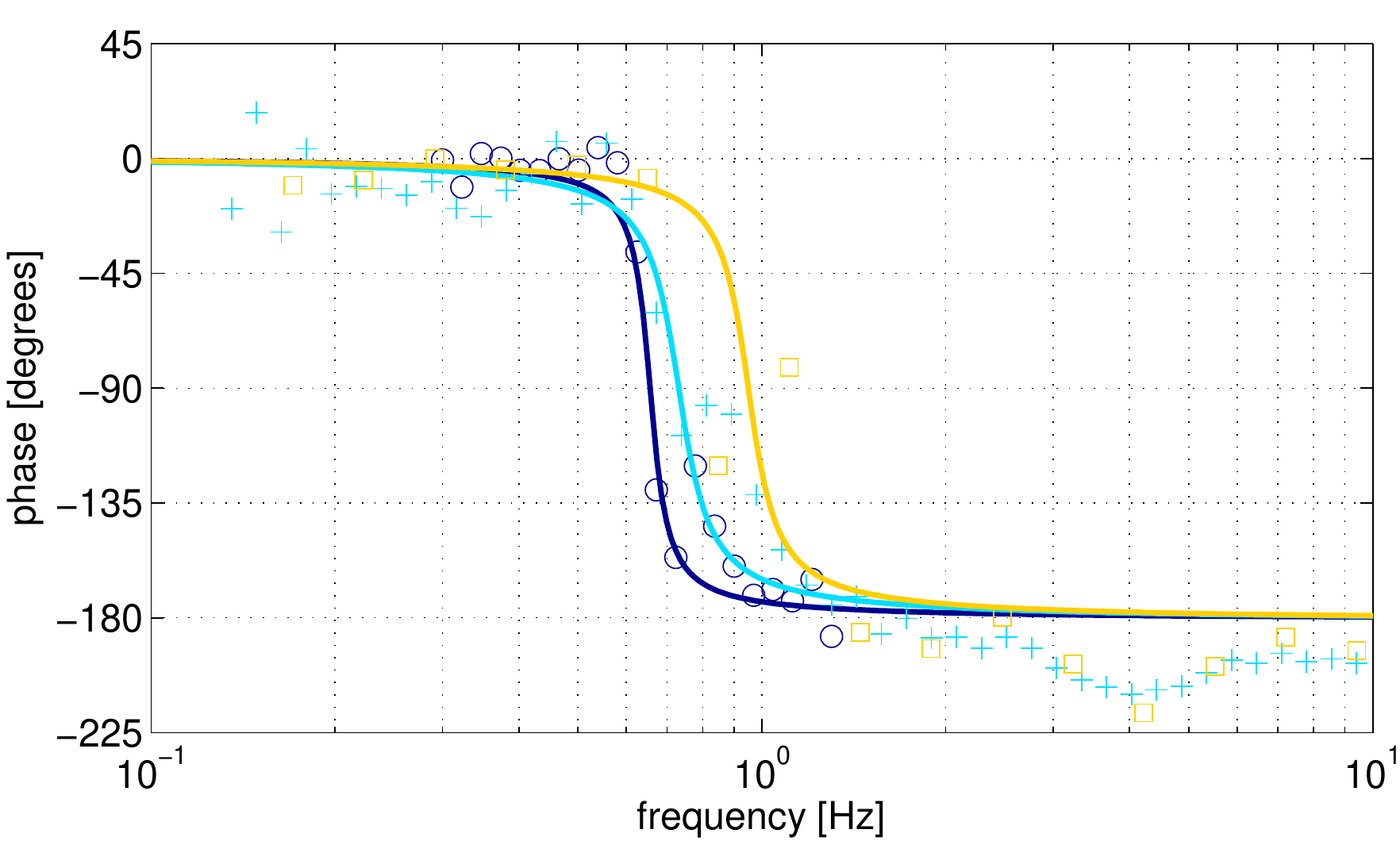}}
\caption{Measured opto-mechanical transfer functions at different
  powers for the A) soft and B) hard degrees of freedom. The resonant
  frequency increases with power for the hard mode and decreases with
  power for the soft mode, which eventually becomes unstable. $P_{\rm
    circ} = 9$ kW was a typical operating power for Initial LIGO and
  $P_{\rm circ} = 40$ kW is the highest of powers reached for Enhanced
  LIGO.  Solid curves indicate fits to the measured data points.  }
\label{fig:hardsoftTF}
\end{figure*}

We directly measured the soft and hard modes for several different
powers, as presented in Figure~\ref{fig:hardsoftTF}, where solid
curves indicate fits to the data to the Sidles-Sigg model. By
injecting an excitation into the control leg of the servo loop and
taking the transfer function from the torque input to the resulting
angle as measured by the WFS (in the radiation pressure eigenbasis),
we reproduce the transfer function of the opto-mechanical plant,
independent of the control system. The measurement points are
highlighted in Figure~\ref{fig:block}. Note that for the measurement
to be independent of the control system, the control system must be
perfectly diagonalized so that it acts as a collection of single-input
single-output loops. Although we periodically tuned the input matrix
to keep the system diagonal, some cross-coupling is expected.

The circulating powers listed in the figure legend are calculated as
follows:
\begin{equation}
P_{\mathrm{circ}} = P_{\mathrm{in}} \epsilon g_{\mathrm{PRC}} T_{\mathrm{BS}} g_{\phi}
\end{equation} 
where $\epsilon=0.7$ is the optical efficiency of the optics between
the laser and interferometer, $g_{\mathrm{PRC}}=35$ is the power gain
of the power recycling cavity, $T_{\mathrm{BS}}=0.5$ is the
transmission of the BS, $g_{\phi}=137$ is power gain of the
Fabry-Perot arms, and $P_{\mathrm{in}}$ is the measured input
power. Our error in the estimation of circulating power is $\pm 20\%$.

Figure~\ref{fig:hardsoftTF}A shows the measurements of the soft
opto-mechanical transfer function for four different circulating
powers. As power increases from 1.7 to 10\,kW, we observe a decrease in
the soft mode resonant frequency from 0.6 to 0.4\,Hz. As the
circulating power is increased to 17\,kW and beyond, the resonance
disappears as expected: the plant has become statically
unstable. Likewise, as shown in Figure~\ref{fig:hardsoftTF}B, we
measured the anticipated increase in resonant frequency as a function
of power for the hard mode. The resonant frequency increases from 0.66
to 0.95\,Hz as the power increases from 1.7 to 17\,kW. We confirm that
the plant for which controls must be designed is no longer that of a
pendulum with a resonance at 0.6\,Hz (pitch) or 0.5\,Hz (yaw), but
that of the soft and hard opto-mechanical system.

\subsection{High laser power angular controls}
We describe an alignment scheme for controlling the mirrors with
radiation pressure dominated angular dynamics which makes use of the
elegant implication of the purely geometric description of a cavity
mode (see Eq.~\ref{eq:kSH}): the opto-mechanical degrees of freedom
remain uncoupled, independent of the circulating power. This is the
core of the ASC work for Enhanced LIGO: to create an input matrix to
rotate the WFS control to the basis of common and differential
opto-mechanical eigenmodes, implement filters in this newly formed
basis, and then increase the gains of only those loops that require
it~\cite{Barsotti2009Modeling}.

The five controlled opto-mechanical degrees of freedom are:
\begin{itemize}
\item differential soft (dSoft) \vspace{-7pt}
\item common soft (cSoft) \vspace{-7pt}
\item differential hard (dHard) \vspace{-7pt}
\item common hard (cHard) \vspace{-7pt}
\item recycling mirror (RM)
\end{itemize}
Each of the two Fabry-Perot cavity arms has a soft and a hard
mode. The use of common/differential degrees of freedom is motivated
by the block-diagonal structure of the sensing matrix in this basis.
The common mode represents motion where the optics of one FP cavity
rotate in the same direction as the other cavity and differential
represents rotations in opposite directions. The recycling mirror is
sensed separately.

The servos which actively control the soft degrees of freedom must be
designed to provide overall stability. This is, in fact, not difficult
because the decreasing resonance naturally moves deeper into the
control bandwidth towards frequencies where gain is higher. At powers
when the resonance disappears, only DC control is
necessary. Counter-intuitively, it is the hard, stable mode, which
poses the greater control challenge. As the resonance of the hard mode
increases with power, it has the potential of making the overall
control loop unstable. The control loop therefore requires a higher
bandwidth than that for the soft mode to both be stable and always
provide damping of the hard mode resonant frequency.

Moreover, the maximum laser power stored inside the LIGO cavities is
reached by progressively increasing the input power. Although the soft
and hard modes remain uncoupled, the servo loops still need to provide
stability over their respective wide ranges of opto-mechanical
transfer functions.

The structure of the angular control loop is to use input and output
matrices before and after the control filters, respectively. The input
matrix realizes the change of basis from the WFS sensors to the
opto-mechanical eigenmodes, and the output matrix changes the basis
once again, from the eigenmodes to individual mirrors. The output
matrix is shown in Table~\ref{table:output}. It is the analytical
basis transformation matrix that diagonalizes the coupled equations of
motion (see Eq.~\ref{eq:kSH}), and is repeated with appropriate sign
changes to form differential and common soft and hard modes of the two
arms. The matrix is arbitrarily normalized so the largest element is
$1$, and $r$ is 0.91 for Livingston and 0.87 for Hanford (a result of
different mirror radii of curvature at each site).  

\begin{table}   
\centering
\caption{WFS output matrix (pitch). The geometry of the arm cavities
  dictates the necessary relative magnitudes of actuation on each
  mirror to create the soft and hard modes. For Livingston $r=0.91$
  and for Hanford $r=0.87$.}
\begin{tabular}{l l l l l l}
\hline 
dSoft & dHard & cSoft & cHard & RM & \\
\hline 
1 & r & 1 & r & 0 & ETMX\\
-1 & -r & 1 & r & 0 & ETMY \\
r & -1 & r & -1 & 0 & ITMX\\
-r & 1 & r & -1 & 0 & ITMY\\
 0 & 0 & 0 & 0 & 1 & RM\\
\hline
\end{tabular}
\label{table:output}
\end{table}

The input matrix is determined experimentally by measuring its
inverse, a sensing matrix which details what combinations of WFS sense
specifically the hard mode or the soft mode. A calibrated sensing
matrix for the radiation pressure eigenbasis as taken during a 17\,kW 
lock is shown in Table~\ref{table:sensing}. Rows represent hard/soft
eigenmode excitation and columns are the WFS signals. Before inverting
the sensing matrix to create the input matrix, the smallest of the
elements (which are equivalent to the elements for which optical gain
is expected to be weak), are set to zero. The elements that remain are
highlighted by boxes. Note that the sensing matrix is in fact composed
of two sub-matrices: one for the differential degrees of freedom, and
one for the common degrees of freedom. Also, due to geometric reasons,
WFS1Q has a particularly strong signal compared to the other WFS which
allows us to provide more control to the dSoft mode than to the other
modes.

\begin{table}
\centering
\caption{Sensing matrix in units of [V/\microrad] (pitch).  All
  elements measured at 9.7\,Hz in the closed-loop system, but with the
  feedback at 9.7\,Hz notched out. Numbers in \textcolor{gray}{gray}
  are the measurement results that have coherence less than 0.9. Boxes
  highlight the elements actually used for computing the control
  servo's input matrix (inverse of sensing matrix); all other elements
  are set to zero.}
\begin{tabular}{l l l l l l}
\hline
WFS1Q & WFS2Q & WFS2I & WFS3I & WFS4I &  \\
\hline
\fbox{2.0}   & \textcolor{gray}{0.03} &\textcolor{gray}{0.06} & \textcolor{gray}{-0.008}  &  \textcolor{gray}{0.01} & dSoft \\
\fbox{0.31}  & \fbox{-0.03} &\textcolor{gray}{-0.04} &  \textcolor{gray}{0.002} & \textcolor{gray}{-0.01} & dHard \\
\textcolor{gray}{0.02} & \textcolor{gray}{-0.01} &  \fbox{0.18} & \fbox{\textcolor{gray}{-0.02}} &  \fbox{\textcolor{gray}{-0.10}} & cSoft \\
\textcolor{gray}{0.17} & \textcolor{gray}{-0.01} & \fbox{-0.21} &  \fbox{\textcolor{gray}{0.007}} & \fbox{-0.12} & cHard \\
\textcolor{gray}{0.09} & \textcolor{gray}{-0.01} & \fbox{-0.21}  &  \fbox{0.04} & \fbox{-0.21} & RM \\
\hline
\end{tabular}
\label{table:sensing}
\end{table}

\section{Results}
\label{sec:results}
In this section we present measurements of the performance of the ASC
system with up to 27\,kW circulating power and demonstrate that the ASC
design meets the LIGO requirements.

\subsection{Open loop gain}
\begin{figure}
\centering
\subfigure{\includegraphics[width=\columnwidth]{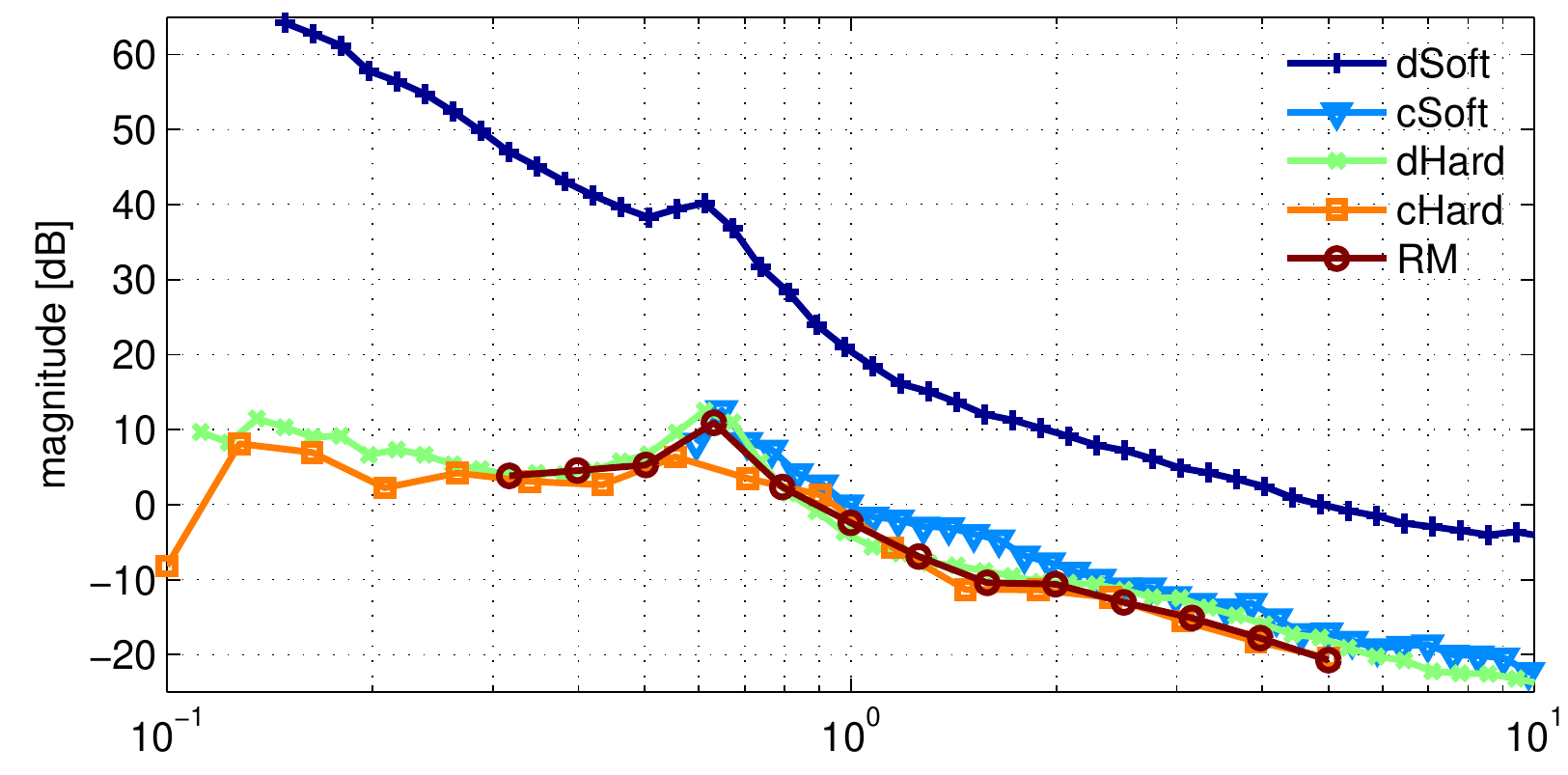}}
\subfigure{\includegraphics[width=\columnwidth]{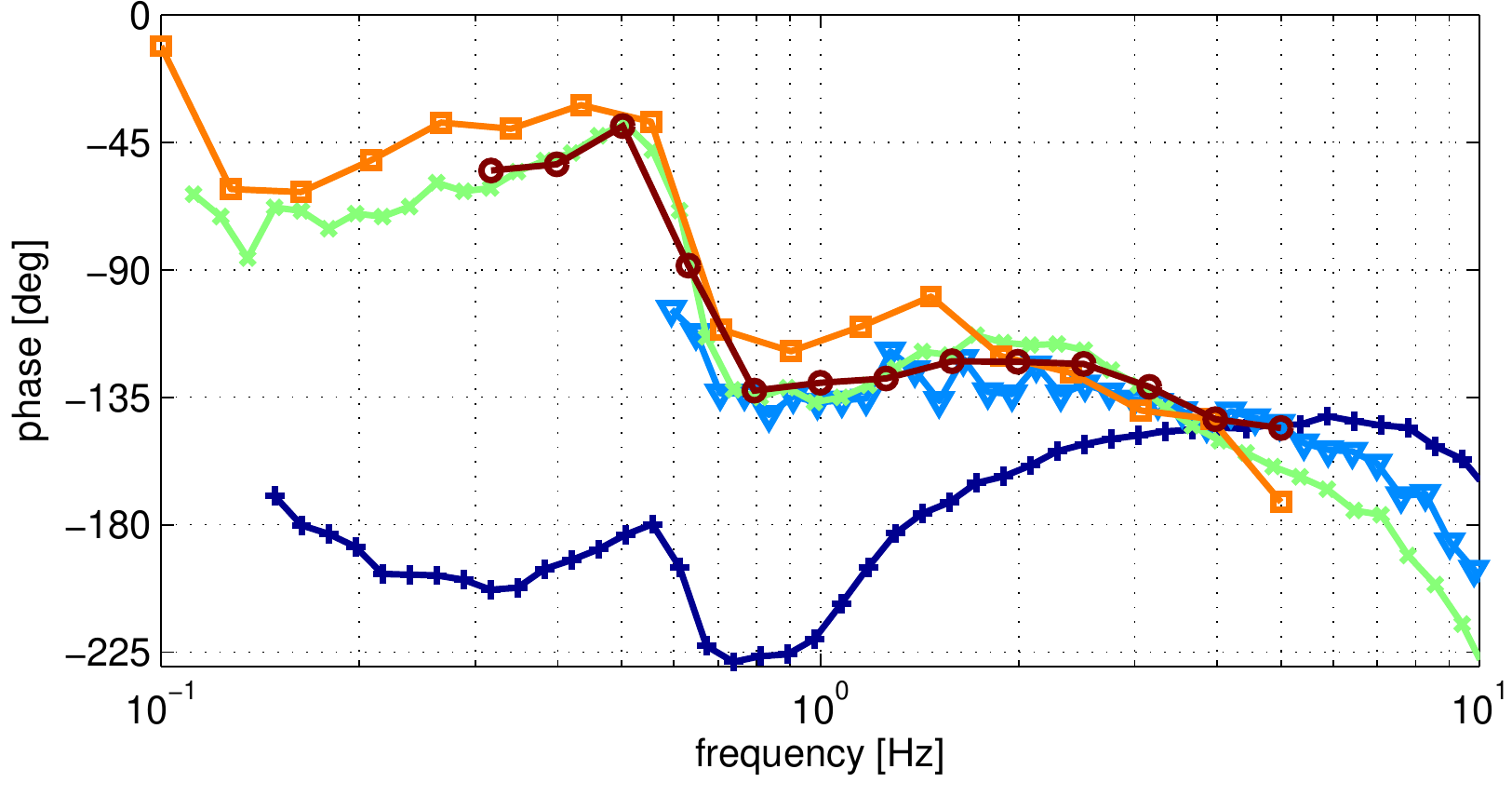}}
\caption{Open loop gains (pitch) of the five WFS loops as measured
  with 10\,kW circulating power. All have a phase margin of 40 to
  $50^\circ$. All UGFs are around 1\,Hz with the exception of the
  dSoft dof whose UGF is 5\,Hz.}
\label{fig:olgs6W}
\end{figure}

The open loop transfer function of each of the WFS loops is the
product of the radiation-pressure-modified pendulum and the control
filters. Figure~\ref{fig:olgs6W} shows the open loop transfer
functions of each of the WFS loops as measured during a 10.3\,kW lock
with the loops closed. As anticipated from the large dSoft signal seen
by WFS1 in the sensing matrix measurement (Table~\ref{table:sensing}),
that is the mode for which we can and do provide the strongest
suppression. In order to achieve this much suppression, it is
necessary to make the feedback loop conditionally stable.  As shown
here, the dSoft unity gain frequency (UGF) is at 5\,Hz. All of the
other degrees of freedom have UGFs of $\sim$\,1\,Hz and are designed
to be unconditionally stable.

The dSoft loop could have a higher gain compared to the other loops
because it caused no harm in strain sensitivity above 60\,Hz, as is
presented later in Section~\ref{sec:noisebudget}. The UGFs of all
other loops were selected as a necessary minimum.

\subsection{Residual beam spot and mirror motion}
\begin{figure}
\centering
\subfigure{\includegraphics[width=0.5\columnwidth]{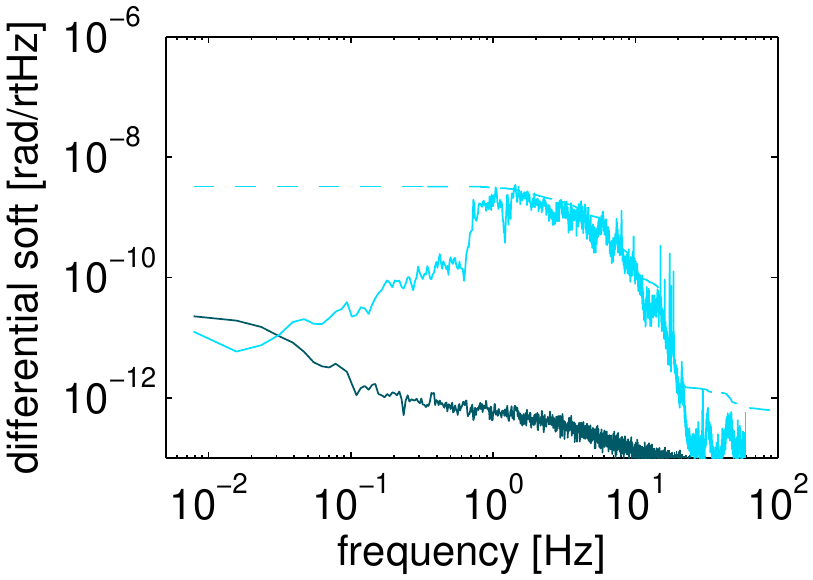}}\subfigure{\includegraphics[width=0.5\columnwidth]{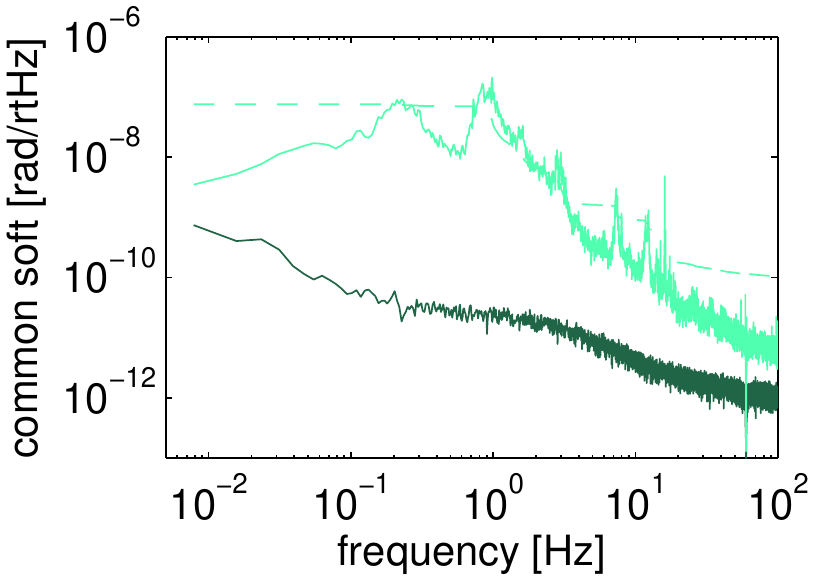}}
\subfigure{\includegraphics[width=0.5\columnwidth]{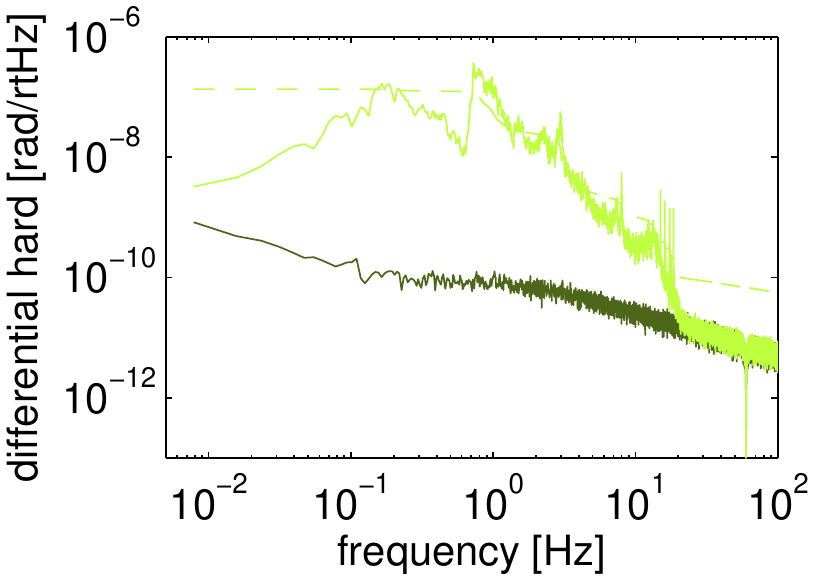}}\subfigure{\includegraphics[width=0.5\columnwidth]{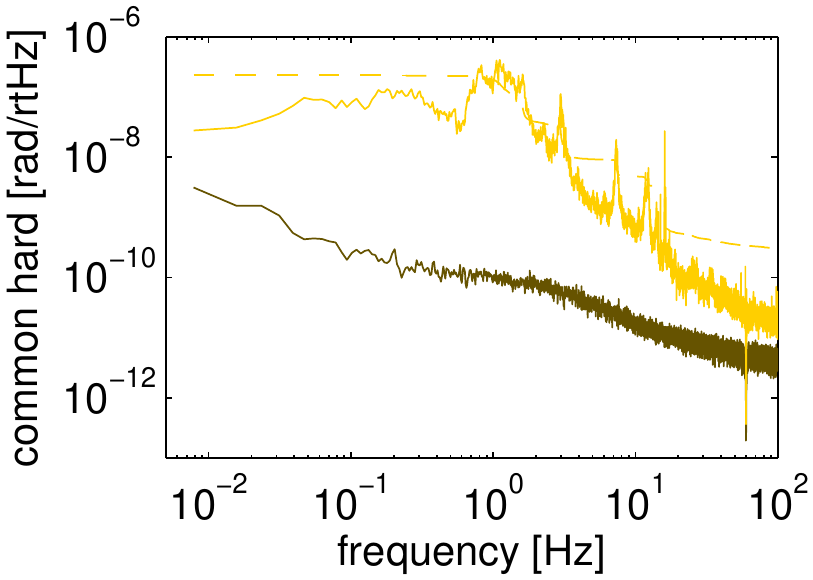}}
\subfigure{\includegraphics[width=0.5\columnwidth]{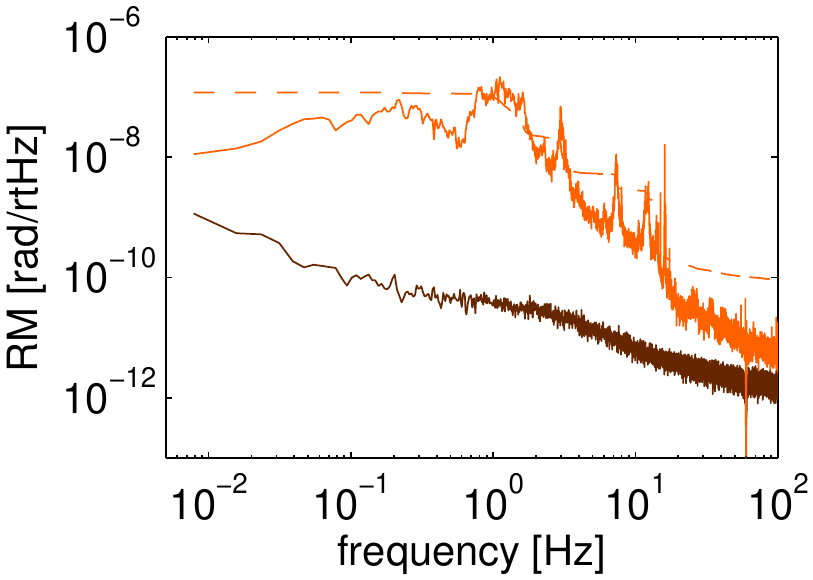}}
\caption{Residual motion of the opto-mechanical degrees of freedom
  during a 17 kW lock. Dashed lines are the rms integrated from the
  right; note that the rms is dominated by the approximately 1\,Hz
  pendular motion. Sensor noise represented in the opto-mechanical
  eigenbasis is also shown.}
\label{fig:FOM}
\end{figure}

Figure~\ref{fig:FOM} shows spectra of the control signal and residual
angular motion in each of the eigenbasis degrees of freedom during a
17\,kW lock. The typical residual rms angular motion is $10^{-7}$
rad/$\sqrt{\mathrm{Hz}}$.

Above 20--25\,Hz, the WFS signals do not represent true angular motion
but instead are limited by a combination of optical shot noise,
photodetector electronics noise and acoustic noise. Unless
sufficiently filtered, the control signal derived from frequencies in
this band will increase the mirror motion.  The resulting need for
low-pass filters limits the achievable bandwidth of the loops.
Because reducing the design UGF allows us to reduce the corner
frequency of the (steep) low pass filters, the reduction of noise in
the GW band is inversely proportional to the UGF raised to the third
or even fourth power. 

\begin{figure}
\centering
\includegraphics[width=\columnwidth]{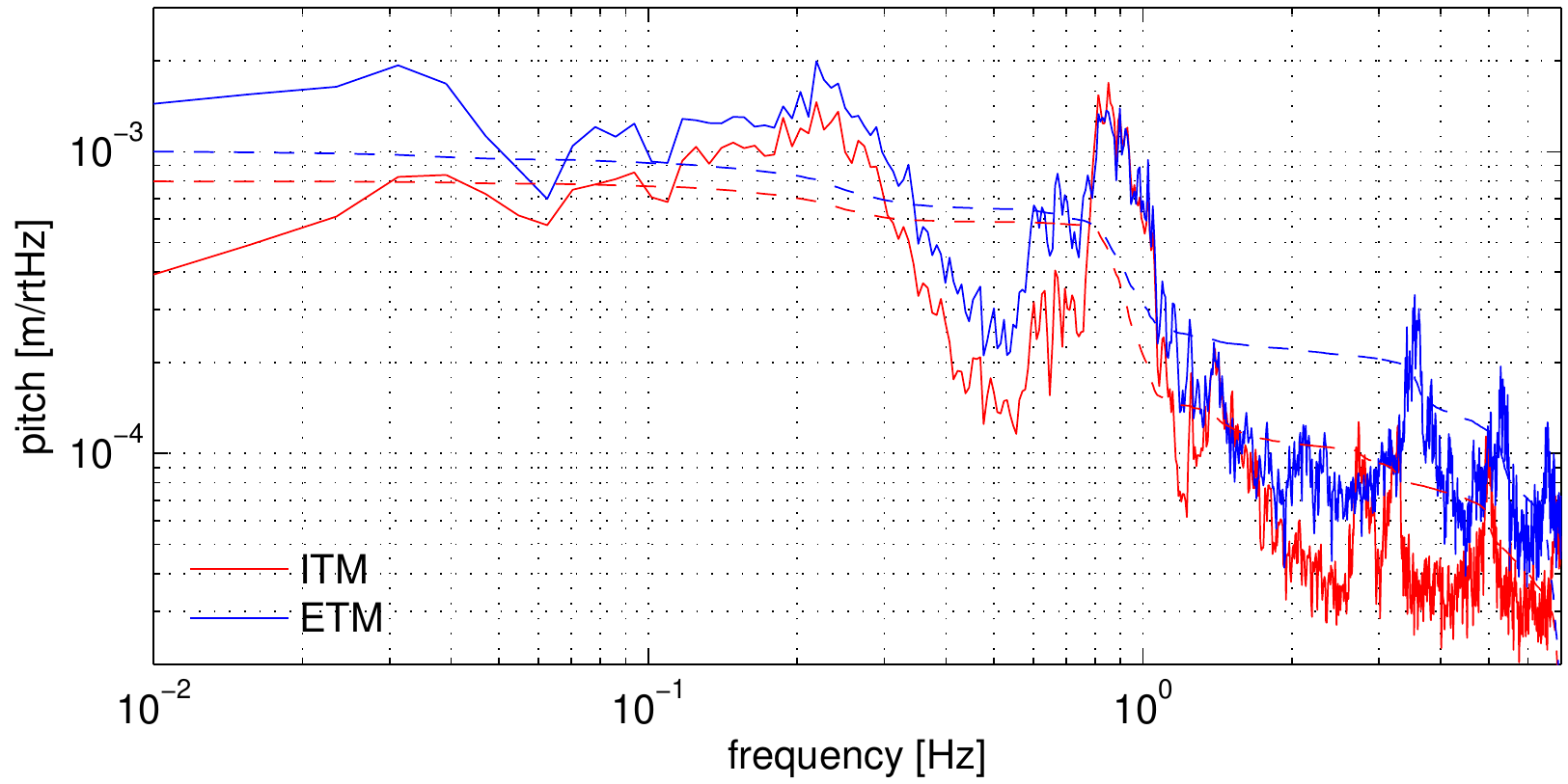}
\caption{Beam spot motion (pitch) on the ITM and ETM mirrors during a
  27\,kW lock at night. Dashed lines are the integrated spectral
  density. For both pitch and yaw, the rms beam spot motion is 1\,mm
  on the ETMs and 0.8\,mm on the ITMs.}
\label{fig:bsm}
\end{figure}

The residual beam spot motion on the test masses is shown in
Figure~\ref{fig:bsm}. The rms beam spot motion on the ETMs is 1\,mm
and on the ITMs it is 0.8\,mm. These measurements are acquired from
the pitch and yaw signals of the QPDs in transmission of the ETMs and
the pitch and yaw DC signal from WFS2 for the ITMs. The magnitudes of
the beam spot motion and the residual mirror motion are
consistent. For example, for $10^{-7}$ rad of soft or hard mode motion
in one arm, we expect the maximum cavity tilt and displacement to be
0.1\,\microrad and 1\,mm, respectively.

\subsection{Angle-to-length coupling (noise performance)}
\label{sec:noisebudget}
One of the most important figures of merit for the control system is
how much noise it contributes to the GW strain signal. As described in
Section~\ref{sec:alignment}, the dominant way in which angular motion
creates a change in cavity length is the convolution of beam spot
motion with angular mirror motion. Ideally, we want the length
displacement due to this coupling to be an order of magnitude below
the desired displacement sensitivity.

The effective transfer function magnitude of the angle-to-length
effects can be estimated with a broadband noise injection that
amplifies the mirror motion. This non-linear technique is necessary
because the linear coupling of torque to cavity length is minimized by
periodically balancing the mirror actuators. Due to the
near-elimination of the linear coupling, the remaining dominant
angle-to-length process (refer to Eq.~\ref{eq:A2L}) has a coupling
coefficient of mean $0$ and the traditional coherent transfer function
measurement would therefore also yield $0$. To arrive at an estimate
of the magnitude of the remaining time-dependent angle-to-length
coupling, the broadband excitation must be averaged over some time. We
injected a 40 to 110\,Hz broadband excitation into the error point
after the input matrix and computed a transfer function between the
hard/soft eigenmode error point and the GW signal. The transfer
function may be multiplied by an ASC signal at any time to estimate a
noise budget.

\begin{figure}
\centering
\subfigure[]{\includegraphics[width=1.0\columnwidth]{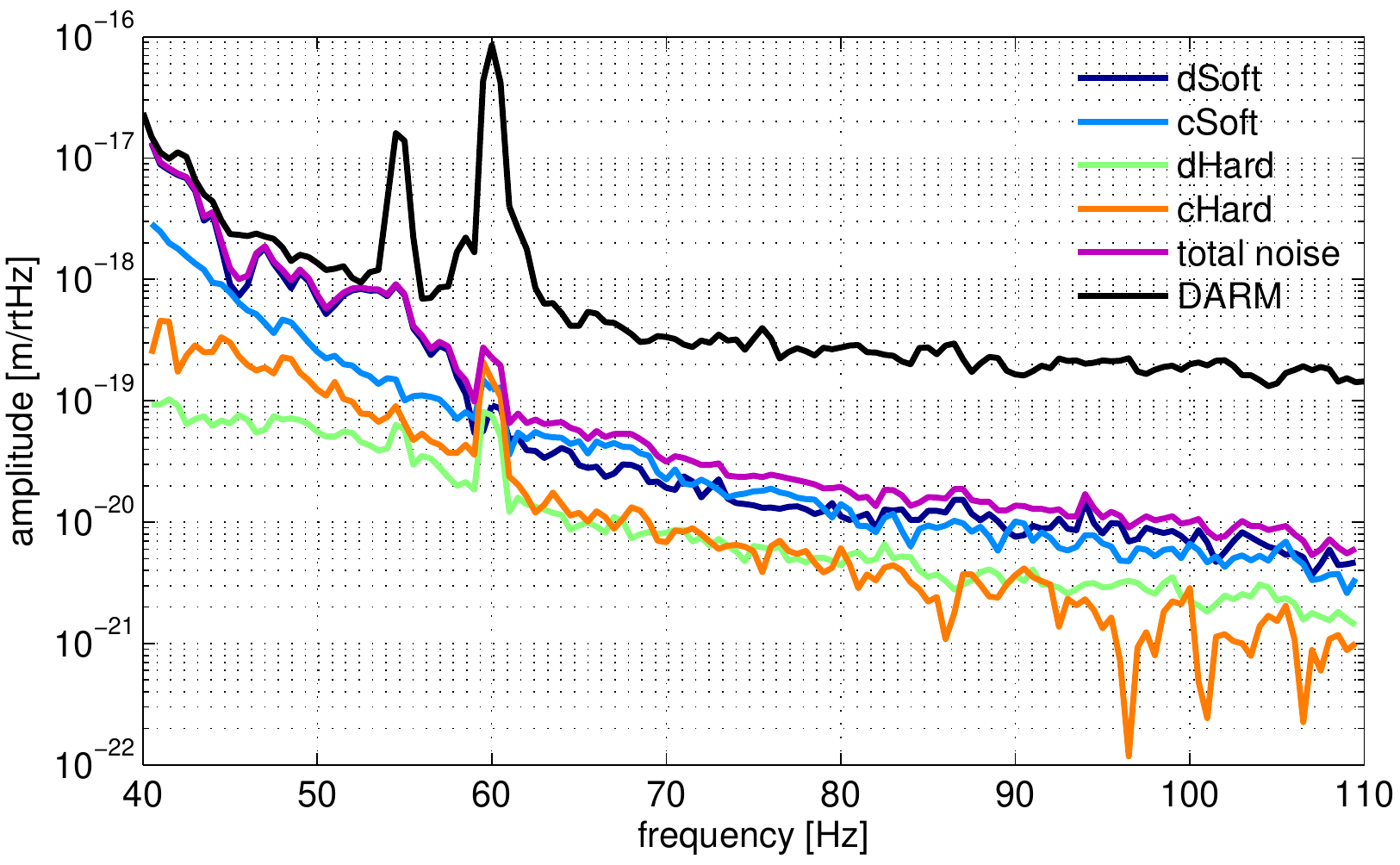}}
\subfigure[]{\includegraphics[width=1.0\columnwidth]{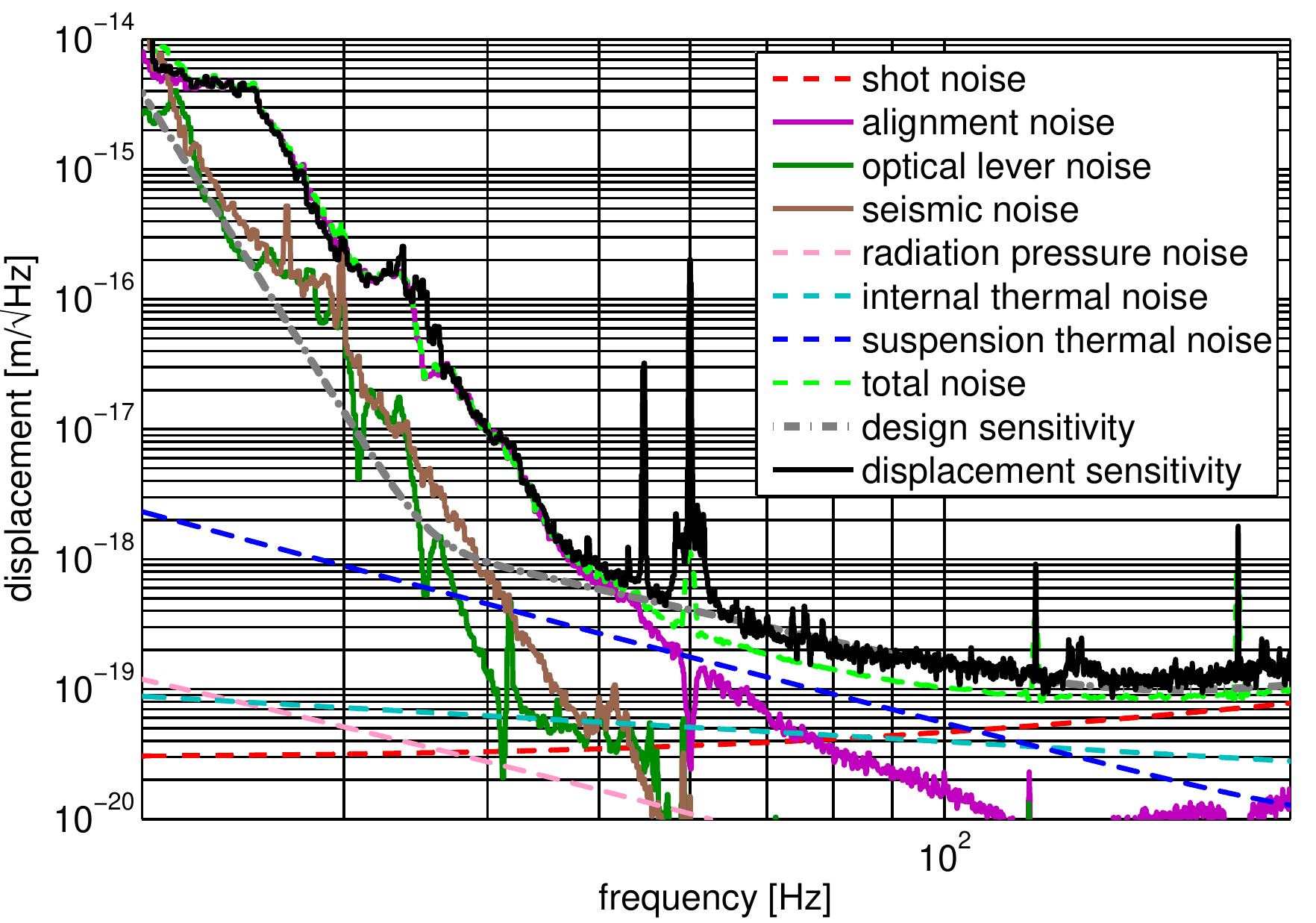}}
\caption{Displacement sensitivity noise budgets during locks with
  24\,kW circulating power. A) Break down of the noise budget of the
  alignment feedback (pitch) dofs to displacement sensitivity. The two
  soft dofs contribute more than the hard dofs. The RM alignment
  feedback is not shown because its contribution is insignificant. B)
  Noise budget of interferometer displacement sensitivity, showing
  several key noise sources. Angular control limits the sensitivity up
  to 55\,Hz. Pitch and yaw contributions are added in quadrature
  because they are de-coupled.}
\label{fig:NB}
\end{figure}

The WFS noise budget in the eigenmode basis for pitch at a time when
the interferometer was locked with 24\,kW power is shown in
Figure~\ref{fig:NB}A. Each degree of freedom's contribution of control
noise to displacement sensitivity is the same within about a factor of
two, except for the RM, which is not included in these plots. We were
not able to measure the transfer function for RM motion to
displacement sensitivity because so large of an excitation was
required to see an effect that the interferometer would lose lock. The
soft modes contributes more length noise than the hard modes. 

The ASC is, in fact, the limiting noise source for frequencies up to
55\,Hz and it becomes less and less of a primary noise source as
frequency increases. By 100\,Hz the ASC noise floor is a factor of 10
below displacement sensitivity. The specific structure of the noise
contributions, including the apparent notches, is a direct result of
the shape of the control filters. Imperfections in the estimate of
displacement noise below 50\,Hz arise because the transfer function is
not perfectly stable in time.

Figure~\ref{fig:NB}B shows a broader view of the role of angular
control noise with respect to other primary noise sources. The
alignment noise shown is the quadrature sum of the pitch and yaw
contributions. Measured seismic and optical lever noises are also
shown, in addition to models of thermal, shot, and radiation pressure
noises. In this example, ASC noise hinders the interferometer
sensitivity up to 55\,Hz by about an order of magnitude. At a later
time, steeper and lower frequency low pass control filters were made
(at the expense of reduced stability) to reduce the alignment noise to
a level similar to that of seismic noise.

\section{Advanced LIGO alignment considerations}
\label{sec:aLIGOASC}
The LIGO detectors are currently being upgraded to a configuration
known as Advanced LIGO to achieve up to a factor of ten improvement in
broadband sensitivity~\cite{Harry2010Advanced}. The noise performance
of the angular control scheme in the Advanced LIGO detectors must meet
the most stringent requirements to date, as imposed by the improved
sensitivity and the goal that the displacement noise produced by the
ASC is no greater than 10\% of the design sensitivity. Given the ASC
was a limiting noise source below 55\,Hz in Enhanced LIGO, some
additional steps must be taken to achieve the Advanced LIGO goal. For
instance, in order to mitigate the largely acoustic dominated WFS
noise above 10\,Hz, the WFS will be placed in vacuum for Advanced
LIGO. In addition, the angle-to-length coupling at low frequencies
will be reduced through the use of a seismic feed-forward scheme.

\begin{figure}
\centering
\includegraphics[width=\columnwidth]{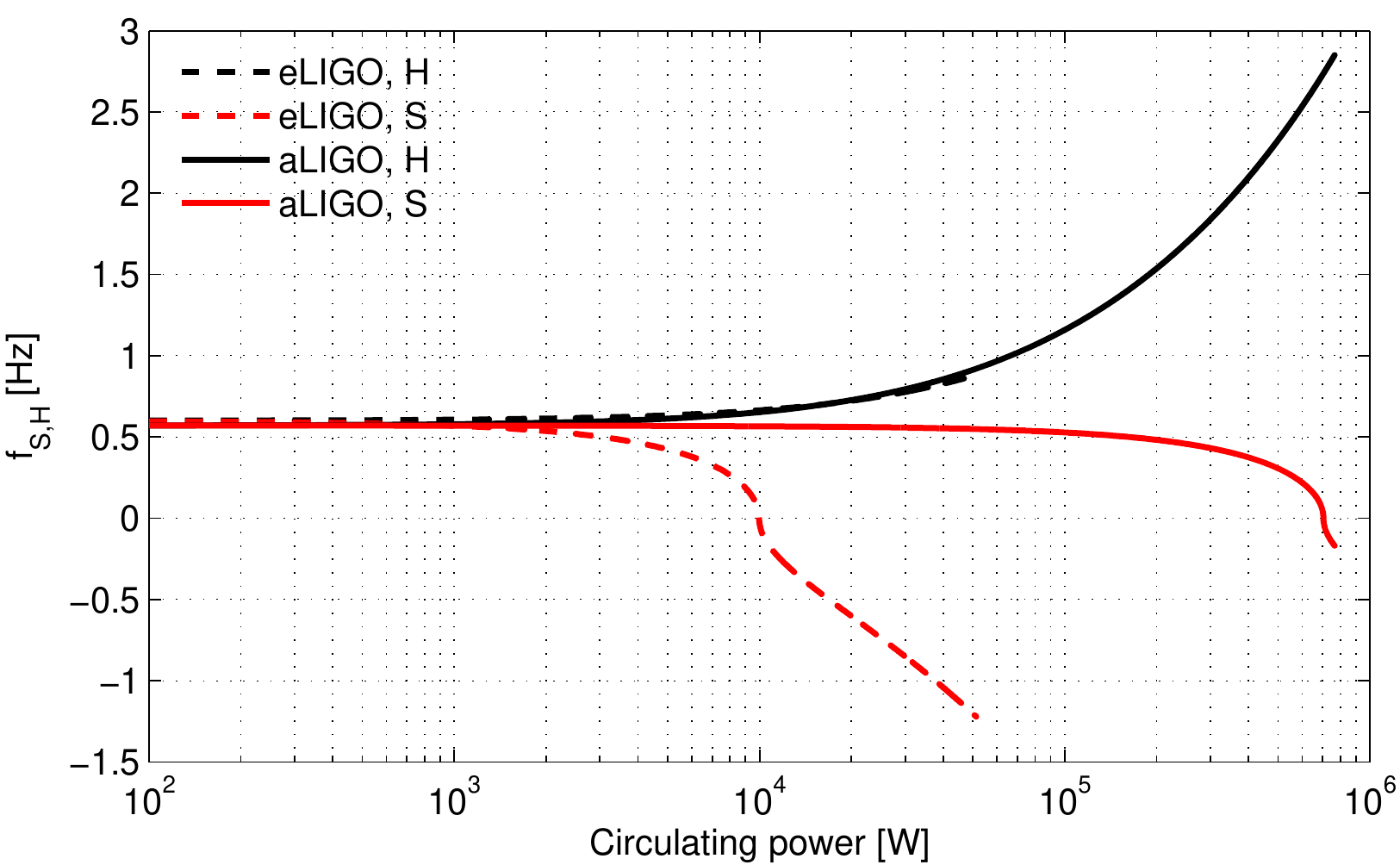}
\caption{Resonant frequency of the opto-mechanical modes for pitch as
  a function of circulating power, comparing Advanced and Enhanced
  LIGO. Only at the highest of Advanced LIGO powers (about 800 kW
  circulating power, or 125 W input power) will the soft mode become
  unstable. Models are plotted up to the highest of their respective
  design powers.}
\label{fig:khardsoft}
\end{figure}

The Sidles-Sigg effect will not be as important in Advanced LIGO
despite the laser power stored in the arm cavities being as high as
800 kW, 20 times higher than in Enhanced LIGO. A number of design
changes have made the impact of radiation pressure less dramatic:
\begin{itemize}
\item four times heavier mirrors (40\,kg instead of 10\,kg);\vspace{-7pt}
\item arm cavity $g$-factor chosen to suppress the soft mode~\cite{Sidles2006Optical}\vspace{-7pt}
\item a larger restoring angular torque due to new multi-stage
  pendulum suspensions
\end{itemize}
Figure~\ref{fig:khardsoft} shows a plot of soft and hard mode
frequency as a function of stored power in the arms for the Enhanced
and Advanced LIGO configurations. It can be seen that although the
hard mode is hardly affected by the Advanced LIGO changes, the new
$g$-factor greatly pushes out the power at which the soft mode becomes
unstable. Nevertheless, the control strategy developed for Enhanced
LIGO gives us confidence that we can control the hard and soft
modes. The Advanced LIGO ASC design detailing the effects of the above
changes to the design presented here is found in
Ref.~\cite{Barsotti2010Alignment}.

\section{Conclusion}
\label{sec:summary}

The Enhanced LIGO interferometer is a complex opto-mechanical system
whose angular mechanics are dominated by radiation pressure effects.
We show that radiation pressure shapes the angular dynamics of the
suspended mirrors and plays an important role in the design of an
angular control system. We implemented and characterized a novel
control scheme to deal with the instabilities that radiation pressure
causes to the angular degrees of freedom of the interferometer,
without compromising the strain sensitivity of the detector. The
alignment control scheme that we describe allowed the LIGO detectors
to operate at their best sensitivity ever, as achieved during the
scientific run S6.  The solution that we demonstrate here is
extensible to the next generation of LIGO detectors, Advanced LIGO,
and is more broadly applicable in systems in which radiation pressure
torques are dominant over mechanical restoring forces.

\section*{Acknowledgements}
We would like to thank D.\,H.~Reitze, D.~Sigg, and Y.~Aso for helpful
discussions. This work was supported by the National Science
Foundation under grant PHY-0757058.  LIGO was constructed by the
California Institute of Technology and Massachusetts Institute of
Technology with funding from the United States National Science
Foundation.  This paper has LIGO document number
\href{https://dcc.ligo.org/LIGO-P1100089/public}{P1100089}.


\end{document}